\documentclass[12pt]{article}
\usepackage{latexsym,amsmath,righttag,chicago,fullpage,graphicx,epsfig}

\newcommand{\mbf}{\mathbf}
\newcommand{\sbf}{\boldsymbol}

\title{\bf Composition Estimation via Shrinkage}
\author{Chong Gu\\{\it Purdue University}}
\date{}

\begin{document}
\maketitle

\begin{abstract}
  In this note, we explore a simple approach to composition
  estimation, using penalized likelihood density estimation on a
  nominal discrete domain.  Practical issues such as smoothing
  parameter selection and the use of prior information are
  investigated in simulations, and a theoretical analysis is
  attempted.  The method has been implemented in a pair of R functions
  for use by practitioners.

\end{abstract}

\section{Introduction}

A composition refers to the proportions of a set of parts that make up
a whole.  For example, the relative abundance of bacterial genera in
some microbiome can be represented by a vector of proportions summing
up to 1.  The analysis of compositional data \cite{aitch:86} found
applications in many fields including metagenomics
(cf. \citeNP{xing:17}).  Our task here is to estimate composition
using empirical data.

\citeN{lhz:20} had excellent discussions of the challenges posed by
composition estimation and the many approaches attempted in the
literature.  Samples of microbiome are probed for the number of
occurrences of bacterial genera, say, and a central issue is the
prevalence of zero counts due to limitations of the probing
technologies.  Zero proportions are not among acceptable answers, and
the challenge is to estimate the proportions associated with zero
counts in some reasoned manner.

Given multiple samples believed to have similar composition patterns,
one may form an empirical composition matrix with columns adding up to
1 but containing entries of zeros, then share information among
columns to produce estimates with desirable properties;
\shortciteN{lhz:20} proposed and illustrated an approach doing just
that, using as estimates some low-rank approximation of the empirical
composition matrix.

In this note, we explore a simple approach to composition estimation.
We take as input a matrix of raw counts, possibly with many $0$'s, and
return a composition matrix of positive entries with the columns
summing up to 1.  The method is actually designed to work with one
column, with information shared by other columns helping to improve
performance; prior information from other sources, if available,
could be equally helpful.

We shall employ penalized likelihood density estimation on a nominal
discrete domain for the task.  The method has long been developed and
used with success in practice, but primarily on continuous domains.
We now take a serious look at how it performs on a discrete domain.
We shall first set the notation and outline the method, then
investigate important practical issues such as smoothing parameter
selection and the effective use of prior information.  A pair of R
functions are available for use by practitioners.  Some theoretical
analysis is also attempted to better understand how the method works.

\section{Method}

Consider a multinomial sample
$\mbf{k}\sim\text{Multinomial}(n;\mbf{p})$, where
$\mbf{p}=(p_{1},\dots,p_{m})^{T}$, $\sum_{y}p_{y}=1$,
$\mbf{k}=(k_{1},\dots,k_{m})^{T}$, $\sum_{y}k_{y}=n$.  Our task is to
estimate $\mbf{p}$.  When $m$ is large and some of the $p_{y}$'s are
small, $k_{y}=0$ is not uncommon, and the simple maximum likelihood
estimate $\hat{p}_{y}=k_{y}/n$ is undesirable.

The estimation of $\mbf{p}$ can be cast as the estimation of
probability density $f(y)$ on domain $\mathcal{Y}=\{1,\dots,m\}$.  An
approach is via the penalized likelihood method, minimizing
\begin{equation}\label{dsty}
  -\frac{1}{n}\sum_{y}\Big\{k_{y}\eta(y)-\log\int_{\mathcal{Y}}e^{\eta(y)}\Big\}
  +\frac{\lambda}{2}J(\eta),
\end{equation}
where $f(y)=e^{\eta(y)}/\int_{\mathcal{Y}}e^{\eta(y)}$, $J(\eta)$ is a
roughness penalty, and the smoothing parameter $\lambda$ controls the
trade-off between goodness-of-fit and smoothness.  See, e.g.,
\citeN{silver:82} and \citeN{gq:93}.  With $y$ nominal, a standard
choice for $J(\eta)$ is $\sum_{y}\big(\eta(y)-\bar{\eta}\big)^{2}$,
where $\bar{\eta}=m^{-1}\sum_{y}\eta(y)$, which ensures invariance
with respect to permutations of elements in $\mathcal{Y}$.  This is
estimation by shrinkage.

Through the specification of $\int_{\mathcal{Y}}g(y)$, one may
incorporate possible prior information concerning $p_{y}$.  With
$\int_{\mathcal{Y}}g(y)=\sum_{y}w_{y}g(y)$, the density $f(y)$ is
relative to the base measure $\{w_{y}\}$ on $\mathcal{Y}$,
$p_{y}=w_{y}e^{\eta(y)}/\sum_{x}w_{x}e^{\eta(x)}$.  Absent prior
information, one simply sets $w_{y}\propto1$, but if there is reason
to suggest that $(p_{y}/\tilde{p}_{y})$'s are near a constant for some
$\tilde{p}_{y}$, say, one may want to use $w_{y}\propto\tilde{p}_{y}$.
A nearly uniform density (relative to a base measure) is easier to
estimate.

With multiple samples
$\mbf{k}_{x}\sim\text{Multinomial}(n_{x};\mbf{p}_{x})$,
$\mbf{p}_{x}=(p_{x,1},\dots,p_{x,m})^{T}$, $\sum_{y}p_{x,y}=1$,
$\mbf{k}_{x}=(k_{x,1},\dots,k_{x,m})^{T}$, $\sum_{y}k_{x,y}=n_{x}$,
where $\mbf{p}_{x}$'s are believed to be close to each other, one may
use the collapsed data $\mbf{k}=\sum_{x}\mbf{k}_{x}$ to estimate a
$\tilde{p}_{y}$ with $w_{y}\propto1$, then use
$w_{y}\propto\tilde{p}_{y}$ for the estimation of individual
$\mbf{p}_{x}$'s.

\section{Density Estimation on Discrete Domain: Practice}

We now look at how (\ref{dsty}) performs on a discrete domain.  Some
theoretical analysis is to be found in Section~5.  Computation is
straightforward.  Our primary interests here are two fold, to check on
the effectiveness of smoothing parameter selection by
cross-validation, and to verify the benefit of using
$w_{y}\propto\tilde{p}_{y}$ when prior information is available.

The cross-validation technique for the selection of $\lambda$ in
(\ref{dsty}) as developed in \citeN{gw:02} proved to be effective on
continuous domains, and plays a central role behind the {\tt ssden}
facility in the R package {\tt gss} \cite{gss:14}; technical details
are to be found in \citeANP{gu:13} (\citeyearNP{gu:13}, Sect. 7.3).
We now explore its performance on a discrete domain via simple
simulation.

Take $m=100$.  We first generate $Z_{y}\sim{N}(0,1)$, then produce 50
sets of $Z_{x,y}=Z_{y}+z_{x,y}$ where $z_{x,y}\sim{N}(0,1/4)$.  One
then has 50 $\mbf{p}_{x}$'s with $p_{x,y}\propto{e}^{Z_{x,y}}$.  Now
split $N=10000=\sum_{x}n_{x}$ total size to these 50 multinomial
distributions in a random but slightly uneven fashion, with $n_{x}$
ranging from 104 to 314.  With the collapsed data of size 10k, one
has a cross-validated estimate $\tilde{p}_{y}$.

For each of the 50 samples, four estimates were calculated, a pair
each with $w_{y}\propto1$ and $w_{y}\propto\tilde{p}_{y}$.  For each
pair, one estimate was with $\lambda$ minimizing the cross-validation
score (as implemented in {\tt ssden}) at $\lambda_{v}$, another with
$\lambda$ minimizing the Kullback-Leibler divergence
$L(\lambda)=\text{KL}(\mbf{p},\hat{\mbf{p}})=\sum_{y}p_{y}\log{p}_{y}/\hat{p}_{y}$
at $\lambda_{o}$; the subscript $x$ is omitted in the formula, but the
50 samples were generated from 50 $\mbf{p}_{x}$'s with generally
different sizes $n_{x}$.

\begin{figure}
\centerline{\includegraphics[height=1.02\linewidth,width=.34\linewidth,angle=270]{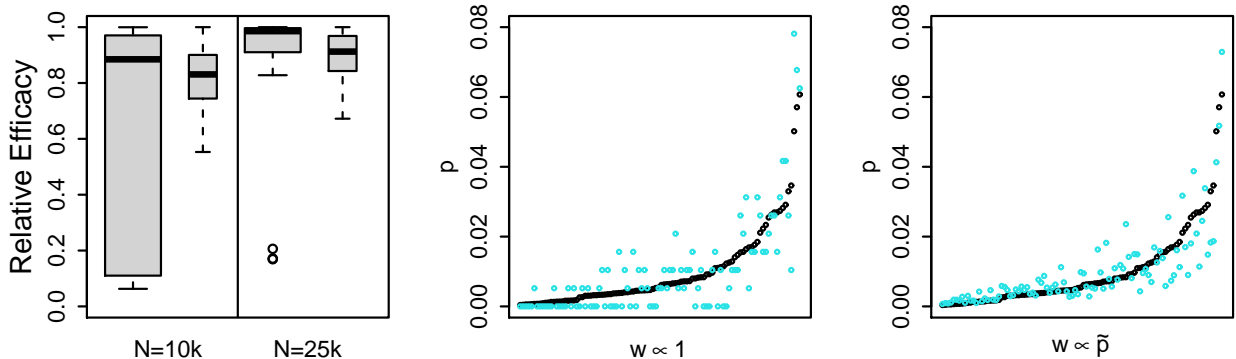}}
\caption{Density Estimation on Discrete Domain: Left: Relative
  efficacy $L(\lambda_{o})/L(\lambda_{v})$ with $w_{y}\propto1$ (wider
  boxes) and $w_{y}\propto\tilde{p}_{y}$ (thinner boxes).  Center and
  Right: $\mbf{p}$ (solid) and cross-validated $\hat{\mbf{p}}$ (faded)
  with $w_{y}\propto1$ (center) and $w_{y}\propto\tilde{p}_{y}$
  (right); $n=192$.}
\label{fig1}
\end{figure}

Ratios of $L(\lambda_{o})/L(\lambda_{v})$ are summarized in the
boxplots in the left half of the left frame in Figure~\ref{fig1}.
The sizes $n_{x}\in[103,314]$ are very low for $m=100$, yet even with
$w_{y}\propto1$, $L(\lambda_{o})$'s were in the range $(0.080,0.257)$
with the median at $0.138$; the bahavior of cross-validation was
dichotomous, succeeding 28 times and failing 22.  The method is
capable, only if one can tune it properly by selecting the right
$\lambda$ in practice.  Using $w_{y}\propto\tilde{p}_{y}$,
$L(\lambda_{o})$'s moved to the range $(0.042,0.100)$ with the median
at $0.064$, and cross-validation performed reliably;
$L(\lambda_{v})$'s had a range $(0.044,0.153)$ with the median at
$0.080$.  Repeating the exercise with a more reasonable total size
$N=25k$, for $n_{x}\in[252,761]$, parallel boxplots are shown in the
right half of the frame; the range of $L(\lambda_{v})$'s for
$w_{y}\propto\tilde{p}_{y}$ are now $(0.033,0.087)$, with the median
at $0.047$.

The cross-validated estimates using a sample of $n=192$ are shown in
the center and right frames in Figure~1, with $p_{y}$'s sorted and
plotted in solid.  The estimates $\hat{p}_{y}$'s in faded, vertically
aligned with the $p_{y}$'s, are invariant to permutations of $y$, but
the sorting cuts down on clutter.  There are 39 $k_{y}=0$ in the
sample.  The $w_{y}\propto1$ fit has
$\text{KL}(\mbf{p},\hat{\mbf{p}})=1.244$ (really bad); the layers of
$\hat{p}_{y}$'s from below in the center frame correspond to
$k_{y}=0,1,\dots$. The $w_{y}\propto\tilde{p}_{y}$ fit has
$\text{KL}(\mbf{p},\hat{\mbf{p}})=0.108$, compared to
$\text{KL}(\mbf{p},\tilde{\mbf{p}})=0.126$.

\section{Miscellaneous}

\subsubsection*{Computation and Software}

On $\mathcal{Y}=\{1,\dots,m\}$, basis functions (i.e., the likes of
unit vectors) are largely independent of each other, so one needs to
entertain $m-1$ coefficients numerically, one each for all but one
$y\in\mathcal{Y}$; the one less is due to a side condition needed on 
$\eta(y)$ to ensure a one-to-one mapping
$f(y)=e^{\eta(y)}/\int_{\mathcal{Y}}e^{\eta(y)}$.  When $m$ is large,
the $O(m^{3})$ execution time could be demanding.

With $w_{y}\propto1$, $\hat{\eta}_{y}$'s associated with $k_{y}=0$ are
all equal by symmetry, as seen in the center frame of
Figure~\ref{fig1}, so one may use that as the baseline, and exclude
bases associated with $k_{y}=0$ to save some computation.

A pair of R functions have been added to the {\tt gss} package to
facilitate the practical use of the proposed method.

One may use {\tt sscomp(x,wt)} to perform density estimation on a
nominal discrete domain, which takes $k_{y}$ in {\tt x}; $k_{y}=0$
does need to be included, as the length of {\tt x} gives $m$.  The
default $w_{y}\propto1$ can be overridden via {\tt wt}.  The function
is simply a stripped down version of {\tt ssden}, tailor-made for use
in the current setting.  It returns the cross-validated estimate
$\hat{\mbf{p}}$ as an $m\times1$ matrix.

With data in a matrix of $k_{x,y}$'s of $m$ rows, one may use {\tt
  sscomp2(x)}, which collapses the columns to estimate
$\tilde{\mbf{p}}$ using {\tt sscomp} with the default $w_{y}\propto1$,
calls {\tt sscomp} with $w_{y}\propto\tilde{p}_{y}$ to estimate
$\mbf{p}_{x}$ for each column, then returns the $\hat{\mbf{p}}_{x}$'s
in a matrix matching the input.

\subsubsection*{Remarks}

Working with the log density $\eta(y)$, one guarantees $p_{y}>0$,
however small they might be.  The domain $\mathcal{Y}$ enters
(\ref{dsty}) via $\int_{\mathcal{Y}}e^{\eta(y)}$, so is part of data
in the setting.

As seen above in the empirical study of Section~3, and also later in
the theoretical analysis of Section~5, $w_{y}$'s mimicking the
``shape'' of $p_{y}$'s allow (\ref{dsty}) to perform better.  This
resembles the mechanism behind importance sampling.  With a single
sample, one may use other sources of prior information in lieu of
$\tilde{\mbf{p}}$, if available.

While $k_{y}$'s should be non-negative integers according to our
multinomial formulation, the algorithm and the R code work with any
non-negative numbers.  The same ratios $k_{y}/n$ represent the same
empirical pattern reflecting $p_{y}$, but a larger $n=\sum_{y}k_{y}$
makes cross-validation to pick a smaller $\lambda$, tilting
$\hat{p}_{y}$ closer to $k_{y}/n$.

\section{Density Estimation on Discrete Domain: Theory}

We now attempt some theoretical analysis concerning density estimation
via (\ref{dsty}) on a nominal discrete domain.  The theory developed in
the literature on continuous domains (\citeNP{silver:82};
\citeNP{coxosu:90}; \citeNP{gq:93}) does not apply here, as conditions
concerning eigenvalues of the quadratic roughness functional
$J(\eta)$ no longer hold.

For a one-to-one mapping
$p_{y}=w_{y}e^{\eta_{y}}/\sum_{x}w_{x}e^{\eta_{x}}$, we impose a side
condition $\sum_{y}\eta_{y}=0$.  It follows that
$J(\eta)=\sum_{y}\eta_{y}^{2}$, and (\ref{dsty}) can be written as
\begin{equation}\label{dsty1}
  -\frac{1}{n}\sum_{y}k_{y}\eta_{y}+\log\sum_{y}w_{y}e^{\eta_{y}}
  +\frac{\lambda}{2}\sum_{y}\eta_{y}^{2}.
\end{equation}

\subsubsection*{Linear Approximation}

Denoting by
$p_{0,y}=w_{y}e^{\eta_{0,y}}/\sum_{x}w_{x}e^{\eta_{0,x}}$ the
true probabilities, substituting $\log\sum_{y}w_{y}e^{\eta_{y}}$ in
(\ref{dsty1}) by its quadratic approximation at $\eta_{0,y}$, and
dropping terms not involving $\eta_{y}$, one has
\begin{multline}\label{quad}
  -\sum_{y}(\tilde{k}_{y}-p_{0,y})\eta_{y}
  +\frac{1}{2}\Big\{\sum_{y}p_{0,y}(\eta_{y}-\eta_{0,y})^{2}
  -\Big(\sum_{y}p_{0,y}(\eta_{y}-\eta_{0,y})\Big)^{2}\Big\}
  +\frac{\lambda}{2}\sum_{y}\eta_{y}^{2}\\
  =-(\tilde{\mbf{k}}-\mbf{p}_{0})^{T}\sbf{\eta}
+\frac{1}{2}(\sbf{\eta}-\sbf{\eta}_{0})^{T}(P_{0}-\mbf{p}_{0}\mbf{p}_{0}^{T})
(\sbf{\eta}-\sbf{\eta}_{0})+\frac{\lambda}{2}\sbf{\eta}^{T}\sbf{\eta},
\end{multline}
where $\tilde{k}_{y}=k_{y}/n$,
$P_{0}=\text{diag}(p_{0,1},\dots,p_{0,m})$; for the quadratic
approximation, write 
$A(\alpha)=\log\sum_{y}w_{y}e^{\alpha(\eta_{y}-\eta_{0,y})+\eta_{0,y}}$,
differentiate with respect to $\alpha$, then
$A(1)\approx{A}(0)+A'(0)+\frac{1}{2}A''(0)$.

We shall first analyze the minimizer of (\ref{quad}), which is linear
in $\tilde{k}_{y}$, then bridge it with the minimizer of
(\ref{dsty1}).  Differentiating (\ref{quad}) with respect to
$\sbf{\eta}$ and setting the gradient to 0, the minimizer
$\tilde{\sbf{\eta}}$ satisfies
\[
(\lambda{I}+P-\mbf{p}_{0}\mbf{p}_{0}^{T})\tilde{\sbf{\eta}}
=(\tilde{\mbf{k}}-\mbf{p}_{0})
+(P_{0}-\mbf{p}_{0}\mbf{p}_{0}^{T})\sbf{\eta}_{0}.
\]
It follows that
\[
\tilde{\sbf{\eta}}-\sbf{\eta}_{0}=(\lambda{I}+P_{0}-\mbf{p}_{0}\mbf{p}_{0}^{T})^{-1}
\big((\tilde{\mbf{k}}-\mbf{p}_{0})-\lambda\sbf{\eta}_{0}\big).
\]

\subsubsection*{Error Bounds}

Write
$B(\alpha)=\sum_{y}h_{y}w_{y}e^{\alpha(\eta_{1,y}-\eta_{0,y})+\eta_{0,y}}
/\sum_{y}w_{y}e^{\alpha(\eta_{1,y}-\eta_{0,y})+\eta_{0,y}}$.  It is clear
that $B(1)-B(0)=\mbf{h}^{T}(\mbf{p}_{1}-\mbf{p}_{0})$, where
$p_{1,y}=w_{y}e^{\eta_{1,y}}/\sum_{x}w_{x}e^{\eta_{1,x}}$.  A Taylor
approximation $B(1)\approx{B}(0)+B'(0)$ yields
$\mbf{h}^{T}(\mbf{p}_{1}-\mbf{p}_{0})\approx\mbf{h}^{T}
(P_{0}-\mbf{p}_{0}\mbf{p}_{0}^{T})(\sbf{\eta}_{1}-\sbf{\eta}_{0})$, so
\begin{equation}\label{skl}
(\mbf{p}_{1}-\mbf{p}_{0})^{T}(\sbf{\eta}_{1}-\sbf{\eta}_{0})
\approx(\sbf{\eta}_{1}-\sbf{\eta}_{0})^{T}(P_{0}-\mbf{p}_{0}\mbf{p}_{0}^{T})
(\sbf{\eta}_{1}-\sbf{\eta}_{0})=V(\sbf{\eta}_{1}-\sbf{\eta}_{0});
\end{equation}
the left-hand-side is $\text{KL}(\mbf{p}_{0},\mbf{p}_{1})
+\text{KL}(\mbf{p}_{1},\mbf{p}_{0})$, the symmetrized
Kullback-Leibler, and the right-hand-side quadratic proxy is a
weighted mean square error of log probability, with $p_{0,y}$'s as the
weights.  Note that $V(\sbf{\eta}+C)=V(\sbf{\eta})$, invariant to the
side condition imposed on $\sbf{\eta}$.  We shall try to bound
\[
(\lambda{J}+V)(\tilde{\sbf{\eta}}-\sbf{\eta}_{0})=
(\tilde{\sbf{\eta}}-\sbf{\eta}_{0})^{T}
(\lambda{I}+P_{0}-\mbf{p}_{0}\mbf{p}_{0}^{T})
(\tilde{\sbf{\eta}}-\sbf{\eta}_{0}).
\]
By Cauchy-Schwartz, one can treat the bias term involving
$\sbf{\eta}_{0}$ and the variance term involving
$(\tilde{\mbf{k}}_{r}-\mbf{p}_{0})$ separately.

For the variance term, note that
$E\big[(\tilde{\mbf{k}}-\mbf{p}_{0})(\tilde{\mbf{k}}-\mbf{p}_{0})^{T}\big]
=n^{-1}(P_{0}-\mbf{p}_{0}\mbf{p}_{0})$, so
\begin{multline*}
E\big[(\tilde{\mbf{k}}-\mbf{p}_{0})^{T}(\lambda{I}+P_{0}-\mbf{p}_{0}\mbf{p}_{0}^{T})^{-1}
(\lambda{I}+P_{0}-\mbf{p}_{0}\mbf{p}_{0}^{T})(\lambda{I}+P_{0}-\mbf{p}_{0}\mbf{p}_{0}^{T})^{-1}
  (\tilde{\mbf{k}}-\mbf{p}_{0})\big]\\
=\frac{1}{n}\text{tr}\big((\lambda{I}+P_{0}-\mbf{p}_{0}\mbf{p}_{0}^{T})^{-1}
(P_{0}-\mbf{p}_{0}\mbf{p}_{0}^{T})\big)
=\frac{1}{n}\sum_{y}\frac{\rho_{y}}{\lambda+\rho_{y}}<\frac{1}{n\lambda},
\end{multline*}
where $\rho_{y}$'s are the eigenvalues of
$P_{0}-\mbf{p}_{0}\mbf{p}_{0}^{T}$, $\sum_{y}\rho_{y}<1$.  For
the bias term,
\[
\lambda^{2}\sbf{\eta}_{0}^{T}(\lambda{I}+P_{0}-\mbf{p}_{0}\mbf{p}_{0}^{T})^{-1}
(\lambda{I}+P_{0}-\mbf{p}_{0}\mbf{p}_{0}^{T})(\lambda{I}+P_{0}-\mbf{p}_{0}\mbf{p}_{0}^{T})^{-1}
\sbf{\eta}_{0}\leq\lambda\,\sbf{\eta}_{0}^{T}\sbf{\eta}_{0}.
\]
Putting thing together, one has
\begin{equation}\label{rate}
(\lambda{J}+V)(\tilde{\sbf{\eta}}-\sbf{\eta}_{0})
=O_{p}\big(\lambda\,\sbf{\eta}_{0}^{T}\sbf{\eta}_{0}+(n\lambda)^{-1}\big).
\end{equation}
When $(w_{y}/p_{0,y})$'s are near a constant,
$\sbf{\eta}_{0}^{T}\sbf{\eta}_{0}$ is small.

These bounds may not be sharp, but in case they are, the optimal rate
also depends on how $\sbf{\eta}_{0}^{T}\sbf{\eta}_{0}$ grows with $m$.
If $\sbf{\eta}_{0}^{T}\sbf{\eta}_{0}\asymp{m}$, then the optimal rate
is $(\lambda{J}+V)(\tilde{\sbf{\eta}}-\sbf{\eta}_{0})
=O_{p}\big(\sqrt{m/n}\big)$ achieved at $\lambda\asymp(mn)^{-1/2}$;
$m$ is allowed to grow with $n$, but the growth rate should be slower
than $O(n)$ to guarantee consistency.

\subsubsection*{Approximation Error}

Set $\eta_{y}=\hat{\eta}_{y}+\alpha{h}_{y}$ in (\ref{dsty1}), where
$\hat{\eta}_{y}$'s minimize (\ref{dsty1}) and $h_{y}$'s are arbitrary.
Differentiating with respect to $\alpha$ and setting the derivative at
$\alpha=0$ to 0, one has
\begin{equation}\label{diff1}
  -\frac{1}{n}\sum_{y}k_{y}h_{y}+\sum_{y}\hat{p}_{y}h_{y}
  +\lambda\sum_{y}\hat{\eta}_{y}h_{y}
  =-\tilde{\mbf{k}}^{T}\mbf{h}+\hat{\mbf{p}}^{T}\mbf{h}
  +\lambda\,\hat{\sbf{\eta}}^{T}\mbf{h}=0,
\end{equation}
where $\hat{p}_{y}=w_{y}e^{\hat{\eta}_{y}}/\sum_{x}w_{x}e^{\hat{\eta}_{x}}$.
Setting $\sbf{\eta}=\tilde{\sbf{\eta}}+\alpha\mbf{h}$ in (\ref{quad}),
differentiating with respect to $\alpha$, then setting the derivative
at $\alpha=0$ to 0, one has
\begin{equation}\label{diff2}
  -(\tilde{\mbf{k}}-\mbf{p}_{0})^{T}\mbf{h}+(\tilde{\sbf{\eta}}-\sbf{\eta}_{0})^{T}
  (P_{0}-\mbf{p}_{0}\mbf{p}_{0}^{T})\mbf{h}+\lambda\tilde{\sbf{\eta}}^{T}\mbf{h}=0.
\end{equation}
Subtracting (\ref{diff2}) from (\ref{diff1}) and setting
$\mbf{h}=\hat{\sbf{\eta}}-\tilde{\sbf{\eta}}$, some algebra yields
\begin{multline}\label{final}
\lambda(\hat{\sbf{\eta}}-\tilde{\sbf{\eta}})^{T}(\hat{\sbf{\eta}}-\tilde{\sbf{\eta}})
+(\hat{\mbf{p}}-\tilde{\mbf{p}})^{T}(\hat{\sbf{\eta}}-\tilde{\sbf{\eta}})\\
=(\tilde{\sbf{\eta}}-\sbf{\eta}_{0})^{T}(P_{0}-\mbf{p}_{0}\mbf{p}_{0}^{T})
(\hat{\sbf{\eta}}-\tilde{\sbf{\eta}})
-(\tilde{\mbf{p}}-\mbf{p}_{0})^{T}(\hat{\sbf{\eta}}-\tilde{\sbf{\eta}}).
\end{multline}
Using the mean value theorem in the arguments leading to (\ref{skl}),
\[
(\hat{\mbf{p}}-\tilde{\mbf{p}})^{T}(\hat{\sbf{\eta}}-\tilde{\sbf{\eta}})
=(\hat{\sbf{\eta}}-\tilde{\sbf{\eta}})^{T}(P_{1}-\mbf{p}_{1}\mbf{p}_{1}^{T})
(\hat{\sbf{\eta}}-\tilde{\sbf{\eta}})
\]
where $P_{1}=\text{diag}(p_{1,1},\dots,p_{1,m})$,
$\mbf{p}_{1}=(p_{1,1},\dots,p_{1,m})^{T}$ correspond to a convex
combination $\sbf{\eta}_{1}$ of $\hat{\sbf{\eta}}$ and
$\tilde{\sbf{\eta}}$.  Likewise,
\[
(\tilde{\mbf{p}}-\mbf{p}_{0})^{T}(\hat{\sbf{\eta}}-\tilde{\sbf{\eta}})
=(\tilde{\sbf{\eta}}-\sbf{\eta}_{0})^{T}(P_{2}-\mbf{p}_{2}\mbf{p}_{2}^{T})
(\hat{\sbf{\eta}}-\tilde{\sbf{\eta}}).
\]
Assuming $\mbf{a}^{T}(P_{i}-\mbf{p}_{i}\mbf{p}_{i}^{T})\mbf{b}$,
$i=1,2$ be bounded by multiples of
$\mbf{a}^{T}(P_{0}-\mbf{p}_{0}\mbf{p}_{0}^{T})\mbf{b}$ from below and
above, one has from (\ref{final}), 
\[
(\lambda{J}+c_{1}V)(\hat{\sbf{\eta}}-\tilde{\sbf{\eta}})
\leq{c}_{2}\big|(\tilde{\sbf{\eta}}-\sbf{\eta}_{0})^{T}(P_{0}-\mbf{p}_{0}\mbf{p}_{0}^{T})
(\hat{\sbf{\eta}}-\tilde{\sbf{\eta}})\big|
\]
for some $0<c_{1}, c_{2}<\infty$.  By Cauchy-Schwartz,
$V(\hat{\sbf{\eta}}-\tilde{\sbf{\eta}})\leq
(c_{2}/c_{1})^{2}V(\tilde{\sbf{\eta}}-\sbf{\eta}_{0})$,
and in turn
$\lambda{J}(\hat{\sbf{\eta}}-\tilde{\sbf{\eta}})\leq
{c}_{3}V(\tilde{\sbf{\eta}}-\sbf{\eta}_{0})$
for some $0<c_{3}<\infty$.  Simple manipulation further propagates the
rate of (\ref{rate}) to
$(\lambda{J}+V)(\hat{\sbf{\eta}}-\sbf{\eta}_{0})$.

\bibliographystyle{chicago}
\bibliography{root}

\end{document}